# Existence of Comfortable Team in some Special Social Networks

Lakshmi Prabha S[1] and T.N.Janakiraman[2,*]

[1]Department of Mathematics, National Institute of Technology,
Trichy-620015, Tamil Nadu, India. Email: jaislp111@gmail.com
[2]Department of Mathematics, National Institute of Technology,
Trichy-620015, Tamil Nadu, India. Email: janaki@nitt.edu
*Corresponding author

**Abstract**

Comfortability is one of the important attributes (characteristics) for a successful team work in any organization. It is necessary to find a comfortable and successful team in any given social network. We have introduced "comfortability" as a new SNA index. Comfortable team exists only in some social networks. In this paper, we analyze the existence of comfortable team in product graphs, such as strong product and Lexicographic product of two given graphs.

**Keywords:** Strong Product; Lexicographic Product ; comfortable team; less dispersive set; domination.

**MSC[2010]:** 91D30 ; 05C69 ; 05C76 ; 05C90.

## 1. Introduction

There are many factors, lack of which affect the group or team effectiveness. The team processes will be described in terms of seven characteristics: coordination, communication, cohesion, decision making, conflict management, social relationships and performance feedback. The readers are directed to refer [1] and [2] for further details of characteristics of team and group dynamics. The attribute "comfortability", is also essential for a successful team work. So, we have introduced "COMFORTABILITY" , as a new SNA index in our paper [3].

Since the beginning of Social Network Analysis, Graph Theory has been a very important tool both to represent social structure and to calculate some indexes, which are useful to understand several aspects of the social context under analysis. Some of the existing indexes (measures or metrics) are betweenness, bridge, centrality, flow betweenness centrality, centralization, closeness, clustering coefficient, cohesion, degree, density, eigenvector centrality, path length. Readers are directed to refer Martino et. al. [4] for more details on indexes in SNA.

Let us define the terminologies as follows: The symbol ($\rightarrow$) denotes "represents"

- Graph $\rightarrow$ Social Network (connected)
- Vertex of a graph $\rightarrow$ Person in a social network
- Edge between two vertices of a graph $\rightarrow$ Relationship between two persons in a social network
- Induced sub graph of a graph $\rightarrow$ **Team or Group** of a social network.

Following are some introduction for **basic graph theoretic concepts**. Some basic definitions from Slater et al. [5] are given below.

The graphs considered in this paper are finite, simple, connected and undirected, unless otherwise specified. For a graph $G$, let $V(G)$ (or simply $V$) and $E(G)$ denote its vertex (node) set and edge set respectively. The length of any shortest path between any two vertices $u$ and $v$ of a connected graph $G$ is called the *distance* between $u$ and $v$ and is denoted by $d_G(u,v)$. For a connected graph $G$, the *eccentricity* $e_G(v) = \max\{d_G(u,v): u \in V(G)\}$. The minimum and maximum eccentricities are the *radius* and *diameter* of $G$, denoted by $r(G)$ and $diam(G)$ respectively. A graph $G$ is self-centered if $r(G) = diam(G)$.

We say that $H$ is a *sub graph* of a graph $G$, denoted by $H < G$, if $V(H) \subseteq V(G)$ and $uv \in E(H)$ implies $uv \in E(G)$. If a sub graph $H$ satisfies the added property that for every pair $u$, $v$ of vertices, $uv \in E(H)$ if and only if $uv \in E(G)$, then $H$ is called an *induced sub graph* of $G$. The induced sub graph $H$ of $G$ with $S = V(H)$ is called the sub graph induced by $S$ and is denoted by $<S|G>$ or simply $<S>$.

The concept of domination was introduced by Ore [6]. A set $D \subseteq V(G)$ is called a *dominating set* if every vertex $v$ in $V$ is either an element of $D$ or is adjacent to an element of $D$. The domination number $\gamma(G)$ of a graph $G$ equals the minimum cardinality of a dominating set in $G$.

Sampath Kumar and Walikar [7] defined a *Connected Dominating Set* (CDS) $D$ to be a dominating set $D$, whose induced sub-graph $<D>$ is *connected*. The minimum cardinality of a connected dominating set is the connected domination number $\gamma_c(G)$.

The readers are also directed to refer Slater et al. [5] for further details of basic definitions, not given in this paper.

## 2. Prior Work

We defined the characteristics of a good performing team and mathematically formulated them in our paper [3]. In order to make this paper self-contained, we give some of the definitions and properties from our paper [3] needed for this paper.

**Definition 1 [3]: Less Dispersive Set**: A set $D$ is said to be less dispersive, if $e_{<D>}(v) < e_G(v)$, for every vertex $v \in D$.

**Definition 2 [3]: Less Dispersive Dominating Set:** A set $D$ is said to be a less dispersive dominating set if the set $D$ is dominating, connected and less dispersive. The cardinality of minimum less dispersive dominating set of $G$ is denoted by $\gamma_{comf}(G)$. A set of vertices is said to be a $<\gamma_{comf}\text{-set}>$, if it is a less dispersive dominating set with cardinality $\gamma_{comf}(G)$.

**Definition 3 [3]: Comfortable Team:** A team $<D>$ is said to be a comfortable team if $<D>$ is less dispersive and dominating. Minimum comfortable team is a comfortable team with the condition: $|D|$ is minimum.

**Example 1:** Consider the graph (network) $G$ in Figure 1. Let $D = \{v_2, v_3, v_4, v_5\}$. We can see from the figure $D$ dominates all the vertices in $V-D$. Also, $<D>$ forms the comfortable team of $G$, because $e_{<D>}(v_2) = 3 < 4 = e_G(v_2)$. $\Rightarrow e_{<D>}(v_2) < e_G(v_2)$. Similarly, $e_{<D>}(v_i) < e_G(v_i)$ for every $i = 3,4,5$. Thus, $D$ forms a less dispersive set and hence $<D>$ forms the comfortable team of $G$. This implies that $\gamma_{comf}(P_6) = 4$.

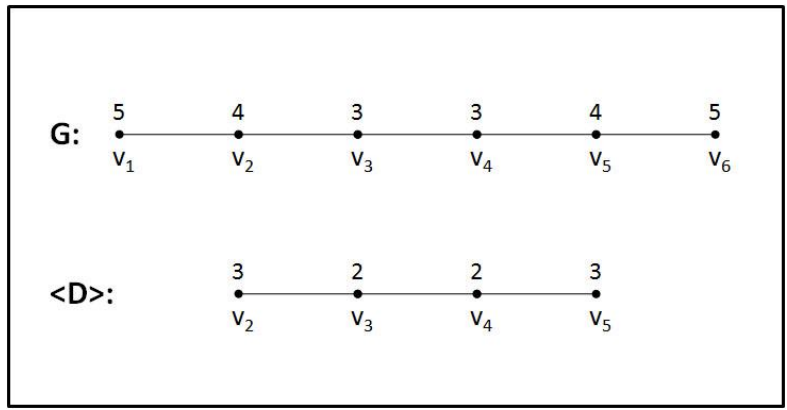

Figure 1: A Network and its Comfortable Team

So, the problem is coined as: Find a team which is dominating, connected and less dispersive.

It is to be noted that there are many graphs which do not have $<\gamma_{comf}\text{-set}>$. So, we must try to avoid such kind of networks for successful team work.

**Example 2:** Consider the graph $G$ in Figure 2. The vertices $v_1$ and $v_4$ dominate all the vertices of $G$. So, with connectedness, we can take $D=\{v_1, v_2, v_3, v_4\}$. The set $D$ dominates $G$, but $D$ is not less dispersive, because, $e_{<D>}(v_1) = 3 = e_G(v_1)$ and $e_{<D>}(v_4) = 3 = e_G(v_4)$. The vertices $v_1$ and $v_4$ maintained the original eccentricity as in $G$. Thus, $e_{<D>}(v) < e_G(v)$, for every vertex $v \in D$ is **not satisfied.** So, $D$ is not less dispersive and hence $<D>$ is not a comfortable team.

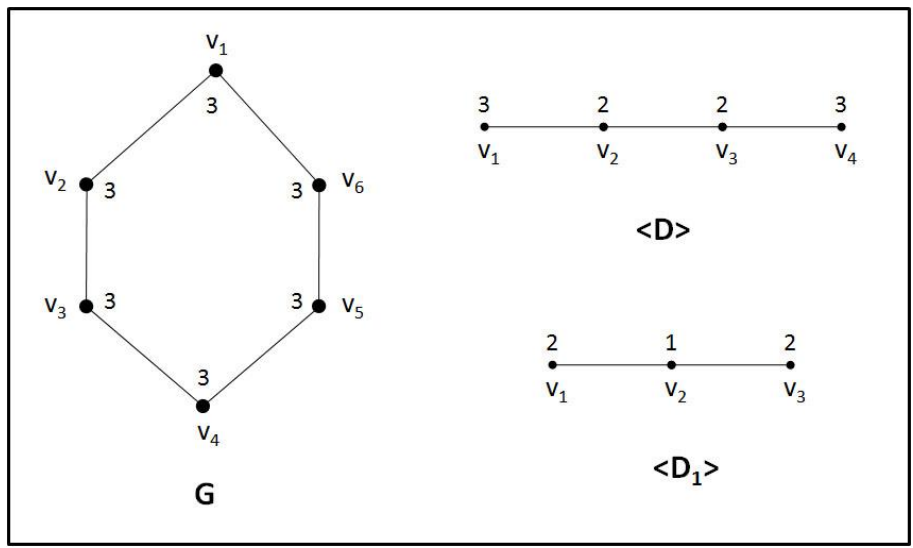

Figure 2. Non-Existence of Comfortable Team

Also, $D_1 = \{v_1, v_2, v_3\}$ forms less dispersive set in $G$, (from Figure 2), but $D_1$ is not dominating. The vertex $v_5$ is left undominated.

From the above discussion, we get,

- the less dispersive set may not be dominating
- the dominating set may not be less dispersive.

So, under one of these two cases, the graph $G$ does not possess comfortable team. It is to be noted that there are infinite families of graphs (social networks) which do not possess comfortable team.

The readers are also directed to refer our paper [3] for further details of comfortable team, not given in this paper.

As comfortable team does not exist in any given graph (social network), we started analyzing the graphs for which comfortable team will exist. It seems to be difficult to characterize the graphs for which comfortable team exists. So, we analyze the existence of comfortable team in some special kind of graphs such as product graphs in this paper.

Let $n$ and $m$ denote the number of vertices in the graphs $G$ and $H$ respectively, that is, $n=|V(G)|$ and $m=|V(H)|$. Let $V(G) = \{u_1, u_2, \ldots, u_n\}$ and $V(H) = \{v_1, v_2, \ldots, v_m\}$.

## 3. Comfortable Team in Strong Product

The strong product of graphs $G$ and $H$ is the graph $G \boxtimes H$, whose vertex set and edge set are

$V(G \boxtimes H) = \{(u_1, v_1) | u_1 \in V(G) \text{ and } v_1 \in V(H)\}$,

$E(G \boxtimes H) = \{(u_1, v_1)(u_2, v_2) | u_1 = u_2 \text{ and } v_1 v_2 \in E(H), \text{ or } v_1 = v_2 \text{ and } u_1 u_2 \in E(G), \text{ or } u_1 u_2 \in E(G) \text{ and } v_1 v_2 \in E(H)\}$.

Figure 3 shows the strong product of two graphs $G$ and $H$.

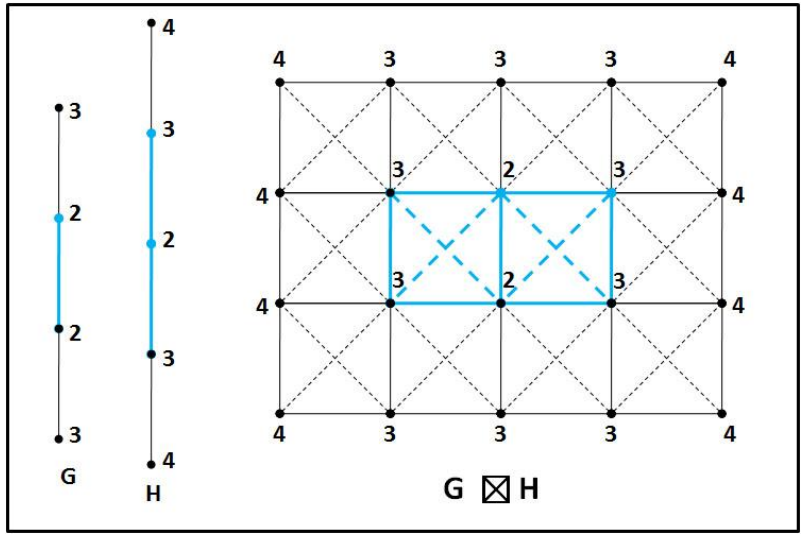

Figure 3: Strong Product of Two Graphs

Next, let us state some existing properties of $G \boxtimes H$.

**Theorem 1** [8]: *Let G and H be graphs. Then for every vertex (u, v) of $G \boxtimes H$, we have*

$e_{G \boxtimes H}((u,v)) = \max \{e_G(u), e_H(v)\}$.

**Theorem 2** [9]: $\gamma(G \boxtimes H) \leq \gamma(G) \cdot \gamma(H)$.

Next, we prove a theorem, which discusses the existence of comfortable team in $G \boxtimes H$.

**Theorem 3:** *If $\gamma_{comf}(G)$ and $\gamma_{comf}(H)$ exists, then $\gamma_{comf}(G \boxtimes H)$ exists and*

$\gamma_{comf}(G \boxtimes H) \leq \gamma_{comf}(G) \cdot \gamma_{comf}(H)$.

*Proof:* Let $S_1$ be a <$\gamma_{comf}$-set> of $G$ and $S_2$ be a <$\gamma_{comf}$-set> of $H$.

**To Prove:** <$S_1 \boxtimes S_2$> forms a comfortable team for $G \boxtimes H$, that is to prove that $S_1 \boxtimes S_2$ dominates $G \boxtimes H$ and $e_{<S_1 \boxtimes S_2>}((u_i, v_j)) < e_{G \boxtimes H}((u_i, v_j))$, for every vertex $(u_i, v_j) \in S_1 \boxtimes S_2$.

**Domination:** As $S_1$ is a <$\gamma_{comf}$-set> of $G$, $S_1$ dominates $G$ (by definition of comfortable team). Similarly, $S_2$ dominates $H$. Also, by Theorem 2, $S_1 \boxtimes S_2$ dominates $G \boxtimes H$.

**Less Dispersiveness:** Let $e_G(u_i) = e_{1i}$ and $e_H(v_j) = e_{2j}$, for every $u_i \in G$ and $v_j \in H$.

$\Rightarrow e_{<S_1>}(u_i) = e_{1i} - k_1$ for some $k_1 \geq 1$ and $e_{<S_2>}(v_j) = e_{2j} - k_2$ for some $k_2 \geq 1$ (by Definition 1), for every $u_i \in S_1$ and $v_j \in S_2$.

By Theorem 1, $e_{G \boxtimes H}((u_i, v_j)) = \max \{e_G(u_i), e_H(v_j)\} = \max \{e_{1i}, e_{2j}\}$ and

$e_{<S_1 \boxtimes S_2>}((u_i, v_j)) = \max \{e_{<S_1>}(u_i), e_{<S_2>}(v_j)\} = \max \{e_{1i} - k_1, e_{2j} - k_2\}$.

**Case (1):** Suppose $e_{<S_1 \boxtimes S_2>}((u_i, v_j)) = e_{1i} - k_1$.

**Sub case(1):** Suppose $e_{G \boxtimes H}((u_i, v_j)) = \max \{e_{1i}, e_{2j}\} = e_{1i}$.

Then, $e_{<S_1 \boxtimes S_2>}((u_i, v_j)) = e_{1i} - k_1 < e_{1i} = e_{G \boxtimes H}((u_i, v_j))$.

**Sub case(2):** Suppose $e_{G \boxtimes H}((u_i, v_j)) = \max \{e_{1i}, e_{2j}\} = e_{2j}$.

This implies that $e_{1i} < e_{2j}$. Then,

$e_{<S_1 \boxtimes S_2>}((u_i, v_j)) = e_{1i} - k_1 < e_{1i} < e_{2j} = e_{G \boxtimes H}((u_i, v_j))$.

As the two sub cases are true for every vertex $(u_i, v_j) \in S_1 \boxtimes S_2$, we get,

$e_{<S_1 \boxtimes S_2>}((u_i, v_j)) < e_{G \boxtimes H}((u_i, v_j))$, for every $(u_i, v_j) \in S_1 \boxtimes S_2$.

**Case (2):** Suppose $e_{<S_1 \boxtimes S_2>}((u_i, v_j)) = e_{2j} - k_2$.

Similar to Case (1), in this case also, we get, $e_{<S_1 \boxtimes S_2>}((u_i, v_j)) < e_{G \boxtimes H}((u_i, v_j))$, for every vertex $(u_i, v_j) \in S_1 \boxtimes S_2$.

Thus, $<S_1 \boxtimes S_2>$ forms a comfortable team for $G \boxtimes H$.

*3.1 Illustration*

Consider the graphs $G$ and $H$ as in the Figure 3 and their strong product. We choose a comfortable team $S_1$ from $G$ (indicated by blue lines in the Figure 3) and a comfortable team $S_2$ from $H$ (indicated by blue lines in the Figure 3). We form the strong product, $S_1 \boxtimes S_2$ (indicated by blue lines in the Figure 3). It can be seen from the Figure 4 that $S_1 \boxtimes S_2$ forms a comfortable team for $G \boxtimes H$.

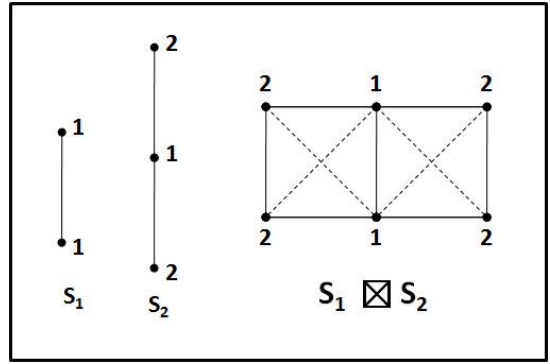

Figure 4: Comfortable Team in Strong Product

## 4. Comfortable Team in Lexicographic Product

The Lexicographic product of graphs $G$ and $H$ is the graph $G \circ H$, whose vertex set and edge set are

$V(G \circ H) = \{(u_1, v_1) | u_1 \in V(G)$ and $v_1 \in V(H)\}$,

$E(G \circ H) = \{(u_1, v_1)(u_2, v_2) | u_1u_2 \in E(G)$ or $u_1 = u_2$ and $v_1v_2 \in E(H)\}$.

Figure 5 shows the Lexicographic product of two graphs $G$ and $H$.

Next, let us state an existing theorem, which will be used to prove our theorem.

**Theorem 4** [9]: $\gamma_c(G \circ H) = \gamma_c(G)$.

From the definition of $E(G \circ H)$, we get the following properties of $G \circ H$. So, we state them without proof:

**Properties:**

1. Any vertex $(u_k, v_j)$ is adjacent to $(u_a, v_j)$, for all $u_a$, which are adjacent to $u_k$ in $G$ and for every $j = 1$ to $m$.
2. If $r(G) = 1$ and $r(H) = 1$, then $r(G \circ H) = 1$ and hence $\gamma_{comf}(G \circ H) = 1$.
3. If $r(G) = 1$ and $r(H) \neq 1$, then $G \circ H$ is 2-self centered and hence $\gamma_{comf}(G \circ H) = 2$. Refer Figure 5, in which $r(G) = 1$.
4. If $r(G) \neq 1$, then $e_{G \circ H}((u_i, v_j)) = e_G(u_i)$, for any vertex $(u_i, v_j) \in G \circ H$. Refer Figure 6.

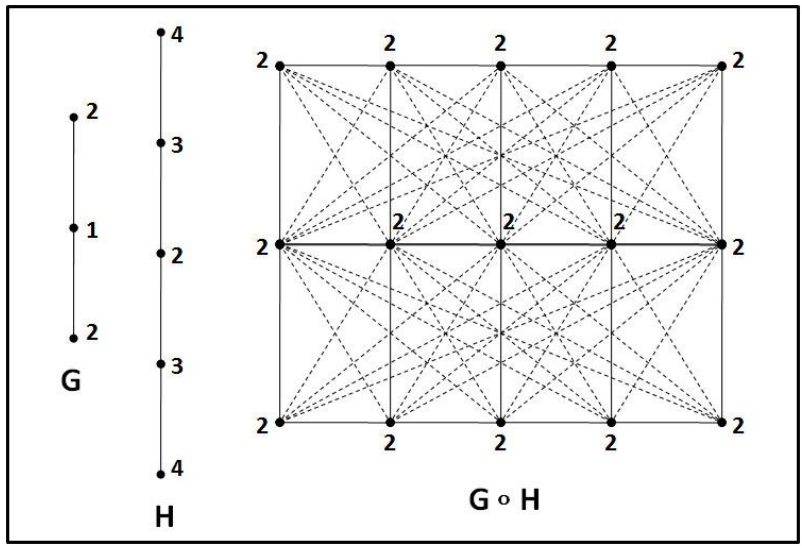

Figure 5: Lexicographic Product of Two Graphs

The following theorem discusses the existence of comfortable team in $G \circ H$.

**Theorem 5:** If $r(G) \neq 1$, then $\gamma_{comf}(G \circ H)$ exists for any $H$ and $\gamma_{comf}(G \circ H) = \gamma_{comf}(G)$.

*Proof:* Let $S$ be a $<\gamma_{comf}$-set$>$ of $G$ and let $T = \{(u_i, v_j): u_i \in S \text{ and } v_j \in H \text{ for any } j=1 \text{ to } m\}$.

This implies that $<T>$ is nothing but $<S>$. So, **$<T>$ and $<S>$ are one and the same**.

**To Prove:** $<T>$ forms a comfortable team for $G \circ H$, that is to prove that $T$ dominates $G \circ H$ and $T$ is less dispersive.

**Domination:** As $S$ is a $<\gamma_{comf}$-set$>$ of $G$, $S$ forms a CDS (connected dominated set) for $G$ (by definition of comfortable team). As $<S>$ and $<T>$ are same and also by Theorem 4, $T$ forms a CDS for $G \circ H$.

**Less Dispersiveness:** As <S> and <T> are same, $e_{<T>}((u_i, v_j)) = e_{<S>}(u_i)$, for every vertex $(u_i, v_j) \in T$. So,

$e_{<T>}((u_i, v_j)) = e_{<S>}(u_i)$

$< e_G(u_i)$ (as $S$ is a less dispersive set for $G$)

$= e_{G \circ H}((u_i, v_j))$ (by Property 4)

Thus, $e_{<T>}((u_i, v_j)) < e_{G \circ H}((u_i, v_j))$, for every vertex $(u_i, v_j) \in T$ and hence $<T>$ forms a comfortable team for $G \circ H$. Also, as $|S| = |T|$, $\gamma_{comf}(G \circ H) = \gamma_{comf}(G)$.

*4.1 Illustration*

Consider the graphs $G$ and $H$ as in Figure 6 and their Lexicographic product $G \circ H$. It can be observed that $S = \{u_2, u_3\}$ (indicated by blue line in the Figure 6) forms a $<\gamma_{comf}$-set$>$ of $G$ and $T = \{(u_2, v_1), (u_3, v_1)\}$ (indicated by blue line in the Figure 6) forms a $<\gamma_{comf}$-set$>$ of $G \circ H$.

Similarly, $\{(u_2, v_j), (u_3, v_j)|$ for any $j= 2$ to $5\}$ will form a $<\gamma_{comf}$ - set$>$ of $G \circ H$. Thus, $\gamma_{comf}(G \circ H) = \gamma_{comf}(G) = 2$ in this case.

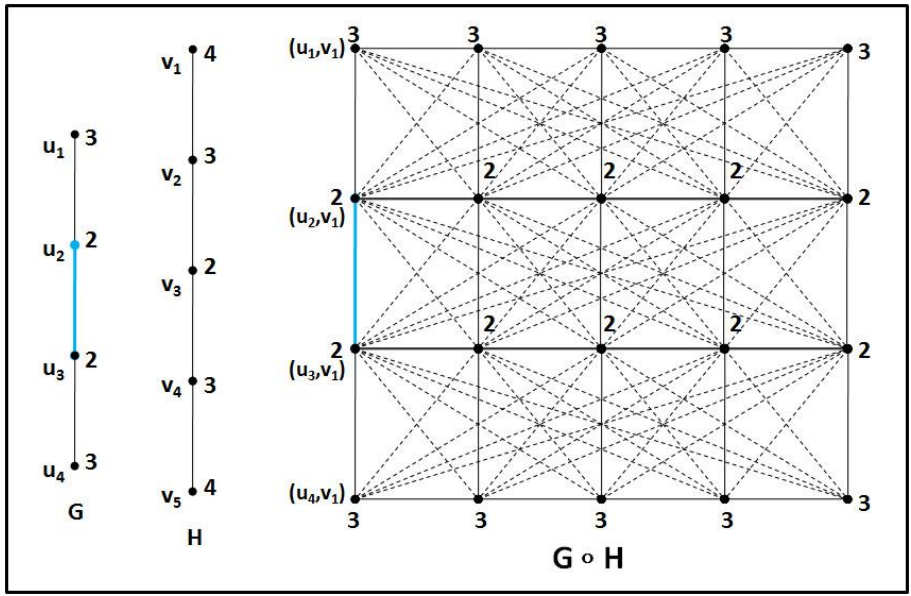

Figure 6: Comfortable Team in Lexicographic Product

## 5. Conclusion

In this paper, the existence of comfortable team are analyzed in some product graphs such as strong product and Lexicographic product of two graphs.

*5.1 Future Directions*

It can be observed from Theorems 3 and 5 that the comfortable team exists in the product of two graphs *G* and *H*, only if both the graphs possess comfortable team. So, further analysis are being carried out for the existence of comfortable team in the product graphs if only one of the graphs possesses comfortable team and if both graphs do not possess comfortable team. Also, analysis are being carried out in other product graphs such as Cartesian products and direct products.

**References**

[1]   S. Mickan and S.Rodger, *Characteristics of effective teams: a literature review*, Australian Health Review, Vol 23, No 3, (2000), 201-208.

[2]   Donelson R.Forsyth, *Group Dynamics*, 3rd ed. Belmont, CA: Wadsworth, (1999).

[3]   Lakshmi Prabha S and T.N.Janakiraman, *Polynomial-time Approximation Algorithm for finding Highly Comfortable Team in any given Social Network*, http://arxiv.org/abs/1405.4534 [CS.DS], [CS.SI], May 2014.

[4]   F.Martino and A.Spoto, *Social Network Analysis: A brief theoretical review and further perspectives in the study of Information Technology*, PsychNology Journal, Volume 4, Number 1, (2006), pp. 53-86.

[5]   T. W. Haynes, S. T. Hedetniemi, P. J. Slater, *Fundamentals of domination in graphs*, Marcel Dekker, New York, (1998).

[6]   O.Ore, *Theory of Graphs*, Amer. Soc. Colloq. Publ. vol. 38. Amer. Math. Soc., Providence, RI, (1962).

[7]   E. Sampathkumar, *H*. B. Walikar, *The connected domination number of a graph*, J.Math.Phys.Sci., 13, (1979), 607-613.

[8]   M. Tavakoli, F. Rahbarnia, and A. R. Ashrafi, *Note on strong product of graphs*, Kragujevac Journal of Mathematics, Volume 37(1) (2013), 187-193.

[9]   R.J.Nowakowski, and D.F.Rall, *Associative graph products and their independence, domination and coloring numbers*, Discussiones Mathematicae. Graph Theory 16 (1996), 53-79.